\documentclass[notitlepage,pre,11pt,tightenlines,%
  nofootinbib,showpacs,showkeys]{revtex4}

\usepackage{amsmath}
\usepackage{amssymb,amsfonts}
\usepackage{bm}
\usepackage[mathcal]{euscript}
\usepackage{graphicx}
\graphicspath{{figs_pre/}}
\usepackage{subfigure}

\newcommand{\ie}{\textit{i.e.}}
\newcommand{\etal}{\textit{et~al.}}
\newcommand{\eg}{\textit{e.g.}}

\newcommand{\mathnotation}[2]{\newcommand{#1}{\ensuremath{#2}}}

\mathnotation{\ldef}{\mathrel{\raisebox{.069ex}{:}\!\!=}}
\mathnotation{\rdef}{\mathrel{=\!\!\raisebox{.069ex}{:}}}
\mathnotation{\nn}{N}				
\mathnotation{\htop}{h}				
\mathnotation{\htopb}{\htop_{\mathrm{braid}}}	
\mathnotation{\htopf}{\htop_{\mathrm{flow}}}	

\begin{document}

\title{Topological Mixing with Ghost Rods}
\author{Emmanuelle Gouillart}
\author{Jean-Luc Thiffeault}
\email{jeanluc@imperial.ac.uk}
\author{Matthew D. Finn}
\affiliation{Department of Mathematics, Imperial College London,
  SW7 2AZ, United Kingdom}
\date{\today}


\begin{abstract}
Topological chaos relies on the periodic motion of obstacles in a
two-dimensional flow in order to form nontrivial braids. This motion generates
exponential stretching of material lines, and hence efficient
mixing. Boyland~\etal~[P. L. Boyland, H. Aref, and M. A. Stremler, J. Fluid
Mech. \textbf{403}, 277 (2000)] have studied a specific periodic motion of
rods that exhibits topological chaos in a viscous fluid. We show that it is
possible to extend their work to cases where the motion of the stirring rods
is topologically trivial by considering the dynamics of special periodic
points that we call \emph{ghost rods}, because they play a similar role to
stirring rods. The ghost rods framework provides a new technique for
quantifying chaos and gives insight into the mechanisms that produce chaos and
mixing. Numerical simulations for Stokes flow support our results.
\end{abstract}

\maketitle

\section{Introduction}
\label{sec:intro}

Low-Reynolds-number mixing devices are widely used in many industrial
applications, such as food engineering and polymer processing.  The study of
chaotic mixing has therefore been an issue of high visibility during the last
two decades.  A first step was taken by Aref~\cite{Aref1984}, who introduced
the notion of \emph{chaotic advection}, meaning that passively advected
particles in a flow with simple Eulerian time-dependence can nonetheless
exhibit very complicated Lagrangian dynamics due to chaos. Chaotic advection
has been demonstrated in many systems since: for a review
see~\cite{Ottino1989} or~\cite{Aref2002}.  However, all the systems considered
had a fixed geometry: even if the boundaries were allowed to move, as for
instance in the journal bearing flow~\cite{Chaiken1986}, the topology of the
fluid region remained fixed.

\begin{figure}
\begin{center}
\includegraphics[width=15cm]{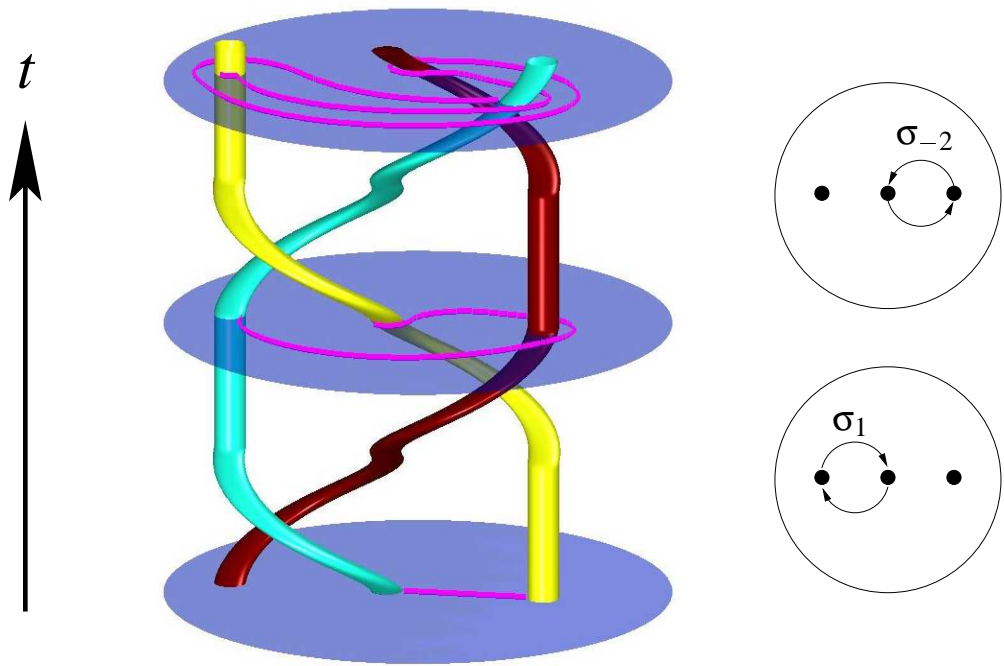}
\end{center}
\caption{The trajectories of the~$\nn$ stirrers define a braid on~$\nn$
strands in a space-time diagram (here~$\nn=3$), with time flowing from bottom
to top.  The periodic movement of the stirrers for the protocol is represented
in the two pictures on the right: first (bottom) the two rods on the left are
interchanged clockwise (we call this operation $\sigma_{1}$), then (top) the
two rods on the right are interchanged anti-clockwise ($\sigma_{-2}$).  Two
sequences of this protocol are drawn on the 3D braid.  This protocol is the
pseudo-Anosov protocol described in~\cite{Boyland2000}. }
\label{fig:trajbraid}
\end{figure}

A new aspect was recently investigated by Boyland~\etal~\cite{Boyland2000}. In
an elegant combination of experimentation and mathematics, the authors
introduced to fluid mechanics the concept of \emph{topological chaos}.  They
studied two different periodic motions of three stirrers in a two-dimensional
circular domain filled with a viscous fluid.  They then used Thurston--Nielsen
(TN) theory~\cite{Thurston1988,Boyland1994} to classify the diffeomorphisms
corresponding to the different stirring protocols.  (The diffeomorphism is a
smooth map that moves the fluid elements forward by one period.)  As the
stirrers moved, the geometry changed in time.  The authors labelled the
protocols using the braid formed by the space-time trajectories of the
stirrers.  In Figure~\ref{fig:trajbraid} we show a space-time plot that
illustrates how the trajectory of the rods can be regarded as a braid, for the
same ``efficient'' braid presented in~\cite{Boyland2000}.

Boyland~\etal\ then used TN theory to determine which stirring protocols
generate \emph{pseudo-Anosov} (pA) diffeomorphisms: a pA diffeomorphism
corresponds roughly to exponential stretching in one direction at every point,
and is thus a good candidate for efficient mixing.  A relevant measure of the
chaoticity of the flow is the maximum rate of stretching of material lines.
In two dimensions this is equivalent to the topological entropy of the
flow~\cite{Newhouse1993}.  In the pA case the braid formed by the stirrer
trajectories gives a lower bound on the topological entropy of the flow
regardless of flow details (\eg\ Reynolds number, compressibility,
\dots)---hence the term \emph{topological chaos}.  A pA braid can only be
formed with three or more rods, since two rods cannot braid around each other
nontrivially, and one rod has nothing to braid with.  In stirring protocols
with pA braids it is thus possible to predict a minimum complexity of the
flow, as opposed to systems that require tuning of parameters to observe
chaos.  Such universality is of course desirable for mixing applications.

In this article we present another aspect of what was described as
\emph{topological kinematics} by Boyland~\etal~\cite{Boyland2003}.  We study
flows with only one stirring rod that have positive topological entropy, even
though the braid traced by the stirrer is trivial.  We apply the topological
theory to these systems by considering the braid formed by periodic orbits of
the flow as well as the stirrer itself.  This allows us to account for the
non-zero topological entropy of the flow. The periodic orbits (which can be
stable or unstable) are created by the movement of the rods, but they are not
the same as the rod trajectories, and in general will have different
periodicity than the rod motions.  We call the periodic points \emph{ghost
rods} because in the context of topological chaos they play the same role as
rods, even though they are just regular fluid particles.  Rather, they are
kinematic rods that act as obstacles to material lines in the flow because of
determinism---a material line cannot cross a fluid trajectory, otherwise the
fluid trajectory must belong to the material line for all times.  In fact, any
fluid trajectory is such a topological obstacle~\cite{Thiffeault2005}, but for
time-periodic systems periodic orbits are the appropriate trajectories to
focus on.  The idea of using periodic orbits to characterise chaos in
two-dimensional systems comes from the study of surface
diffeomorphisms~\cite{Boyland1994}.  In a related vein, periodic orbit
expansions are also used to compute the average of quantities on attractors of
chaotic systems~\cite{Auerbach1987,Cvitanovic1988}.

The study of ghost rods is important because it helps identify the source of
the chaos (and hence good mixing) in a given mixer.  We will show that the
main contribution to the topological entropy in a system usually comes from a
relatively small number of periodic orbits.  This represents a tremendous
reduction in the effective dimensionality of the system, and by focusing on
this reduced set of orbits it will be easier to study and improve mixing
devices. The ghost rods framework thus provides new tools for diagnosing and
measuring mixing~\cite{Thiffeault2005}. In addition, it also gives a new
understanding of the mixing mechanisms as we can consider the mixing to arise
from the braiding of material lines around the ghost rods.

The outline of the paper is as follows.  In Section \ref{sec:braids} we
introduce the mathematical theory for braids and topological chaos.  In
Section~\ref{sec:evidence} we study examples with one rod moving on different
paths.  We show that some periodic orbits braid with the stirrer.  In
Section~\ref{sec:braiding} we show that we can account for an arbitrary
percentage of the observed topological entropy of the flow with such a
braid. The main conclusions and an outlook on future research are presented in
Section \ref{sec:discussion}.

\section{Braids and Dynamical Systems}
\label{sec:braids}

\subsection{Overview of Thurston--Nielsen (TN) Theory}

The mathematical setting for studying braiding in fluids mechanics is centered
on the $\nn$-punctured disk in two dimensions, $R_{\nn}$.  The~$\nn$
punctures, located somewhere in the interior of the disk, represent the
stirring rods.  If the stirrers undergo a prescribed periodic motion, they
return to their initial position at the end of a full cycle.  Naturally, the
rods have dragged along the fluid, which obeys some as yet unspecified
equations (\eg, Stokes, Navier--Stokes, Euler, non-Newtonian equations,
assumption of incompressibility, \dots).  The position of fluid elements is
thus determined by some function~$\Phi(x,t)$.  Since the flow is periodic with
period~$T$,~$\Phi(x,T)$ is a map from~$R_{\nn}$ to itself, and we define
\begin{equation}
  f: R_{\nn} \rightarrow R_{\nn}\, , \qquad f(x)=\Phi(x,T).
\end{equation}
For any realizable fluid motion,~$f$ will be an orientation-preserving
diffeomorphism of~$R_{\nn}$.  The map~$f$ takes every fluid element to its
position after a complete cycle.  It is a diffeomorphism because the physical
fluid flows we consider are differentiable and have a differentiable inverse.
The map~$f$ embodies everything about the fluid movement, and it is thus the
main object to be studied.

But in some sense~$f$ contains too much information: it amounts to a
complete solution of the problem, and there are interesting things we can
say about the general character of~$f$ without necessarily solving for
it.  We are also interested in knowing what characteristics of the
stirrer motion must be reflected in~$f$.  This is where the concept of an
\emph{isotopy class} comes in: two diffeomorphisms are \emph{isotopic} if
they can be continuously deformed into each other.  This is a strong
requirement: continuity means that two nearby fluid elements must remain
close during the deformation, and so they are not allowed to ``go
through'' a rod during the deformation, otherwise they would cease to be
neighbours.  In that sense, isotopy is a topological concept, since it is
sensitive to obstacles in the domain (the rods).  The isotopy class
of~$f$ is then the set of all diffeomorphisms that are isotopic to~$f$.
In fact, an entire class can be represented by just one of its members,
appropriately called the \emph{representative}. Of course, the important
point here is that that not all diffeomorphisms are isotopic to the
identity map.  Note that the overwhelming majority of~$f$ are not allowable
fluid motions (\ie, they cannot arise from the dynamical equations
governing the fluid motion), but any allowable fluid motion must belong
to some isotopy class.

The problem now is to decide what isotopy classes are possible, and what this
means for fluid motion.  Thurston--Nielsen (TN)
theory~\cite{Thurston1988,Boyland1994} guarantees the existence in the isotopy
class of $f$ of a representative map, the TN representative~$f'$, that
belongs to one of three categories: \begin{enumerate}

\item \emph{finite-order}: if~$f'$ is repeated enough times, the resulting
  diffeomorphism is isotopic to the identity (\ie, ${f'}^{m}$ is isotopic to
  the identity for some positive integer $m$);

\item \emph{pseudo-Anosov} (pA): $f'$ stretches fluid elements by a
  factor~$\lambda > 1$, so that repeated application gives exponential
  stretching;

\item \emph{reducible}: $f'$ can be decomposed into components acting on
smaller regions of the previous two types.

\end{enumerate}

A famous example of an Anosov map is Arnold's cat map~\cite{Arnold}. A pA map
is an Anosov map with a finite number of singularities.  The TN theory also
states that the dynamics of a diffeomorphism in a pA isotopy class are at
least as complicated as the dynamics of its pA TN representative, meaning it
has a greater or equal topological entropy.  For a diffeomorphism, the
topological entropy~\htop\ gives a measure of the complexity, \emph{i.e.} the
amount of information which we lose at each iteration of the
map~\cite{Newhouse1993}. It also describes the exponential growth rate of the
number of periodic points as a function of their period. Newhouse and
Pignataro~\cite{Newhouse1993} noticed that~\htop\ also gives the exponential
growth rate for the length of a suitably chosen material line.  In the
numerical simulations described below, we use this fact to compute the
topological entropy of a flow: we consider a small blob in the chaotic region
of the flow and we calculate the growth rate of its contour length.

TN theory tells us about the possible classes of diffeomorphisms that can
arise from the periodic motion of the stirrers.  But one can also consider the
trajectories of the punctures (here the stirrers) in a 3D space-time plot
(Fig.~\ref{fig:trajbraid}), where the vertical is the time axis.  These
trajectories loop around each other and form a physical braid.  The crucial
point is that such a braid constructed from the rod trajectories specifies the
isotopy class of~$f$, no matter the details of the flow.  (See
Refs.~\cite{Boyland1994,Boyland2000,Boyland2003,Birman2004} for further
details.)  As a consequence of the TN theory, the topological entropy of this
braid is a lower bound on the topological entropy of the flow.

Hence, determining the isotopy class of the diffeomorphism~$f$ is
equivalent to studying the braid traced by the stirrer trajectories. This
is a drastic reduction in complexity, because we are free to impose
relatively simple braiding on the stirrers by means of a short sequence
of rods exchanges, whereas the resulting diffeomorphism obtained by
solving the fluid equations can be quite complicated.  In the next
section we introduce the machinery needed to characterize braids.

\subsection{Artin's Braid Group}

Let us introduce now the notation for braids.  The generators of the Artin
braid group on~$\nn$ strands are written $\sigma_{i}$ and $\sigma_{i}^{-1}
\rdef \sigma_{-i}$, which represent the interchange of two adjacent strands at
position~$i$ and~$i+1$.  The interchange occurs in a clockwise fashion for
$\sigma_{i}$ ($i$ goes over $i+1$ along, say, the $y$-axis) and anti-clockwise
for $\sigma_{-i}$ ($i$ goes under $i+1$).  For~$\nn$ strands, there
are~$\nn-1$ generators, so~$i \in \{1,\ldots,\nn-1\}$.  It is thus possible to
keep track of how $\nn$ rods are permuted and of the way they cross by writing
a \emph{braid word} with the ``letters'' $\sigma_{i}$.  We read braid words
from left to right (that is, in~$\sigma_1\sigma_3\,\sigma_{-2}$ the generator
$\sigma_1$ precedes $\sigma_3$ temporally).  For example, the pA stirring
protocol described by Boyland~\etal~\cite{Boyland2000} and shown in
Fig.~\ref{fig:trajbraid} corresponds to $\sigma_{1}\sigma_{-2}$: it consists
of first interchanging the two rods on the left clockwise ($\sigma_{1}$) and
then interchanging the two rods on the right anti-clockwise ($\sigma_{-2}$).
Note that the~$i$ index on~$\sigma_i$ refers to the relative position of a rod
(\eg, second from the left along the~$x$-axis) and does not always label the
same rod.

The generators obey the \emph{presentation} of the braid group,
\begin{subequations}
\begin{alignat}{2}
\sigma_i \sigma_{j} \sigma_i &= \sigma_{j} \sigma_{i} \sigma_{j} \qquad
&\mbox{if} \qquad |i-j| &= 1, \label{eq:121} \\
\sigma_i \sigma_j &= \sigma_j \sigma_i \qquad &\mbox{if} \qquad |i-j| &\ge 2,
\label{eq:comm}
\end{alignat}%
\label{eq:presentation}%
\end{subequations}%
that is, the relations~\eqref{eq:presentation} must be obeyed by the
generators of physical braids, and no other nontrivial relations exist in the
group~\cite{Birman2004}.  The relation~\eqref{eq:121} corresponds physically
to the ``sliding'' of adjacent crossings past each other, and~\eqref{eq:comm}
to the commutation of nonadjacent crossings.

The braid group on~$\nn$ strands has a simple representation in terms
of~$(\nn-1)\times(\nn-1)$ matrices, called the Burau
representation~\cite{Burau1936}.  For three strands ($\nn=3$), the topological
entropy of the braid is obtained from the magnitude of the largest eigenvalue
of the Burau matrix representation of the braid word.  This is the technique
used in Refs.~\cite{Boyland2000,Boyland2003,Vikhansky2003,Finn2003} to compute
topological entropies.  For~$\nn>3$ the Burau representation only gives a
lower bound on the topological entropy of the braid~\cite{Kolev2003}, so a
more powerful algorithm must be used to obtain accurate values.  Here we use
the train-tracks code written by T. Hall~\cite{HallTrain}, an implementation
of the Bestvina--Handel algorithm~\cite{Bestvina1995}.  The train-tracks
algorithm works by computing a graph of the evolution of edges between rods
under the braid operations.  It suffices for our purposes to say that
train-tracks determine the shortest possible length of an ``elastic band''
that remains hooked to the rods during their motion---the minimum stretching
of material lines~\cite{Finn2005}.

\section{Evidence for Ghost Rods}
\label{sec:evidence}

\subsection{Motivation}
\label{sec:trm}

We consider here the advection of a passive scalar in a two-dimensional batch
stirring device containing a viscous fluid that obeys Stokes' equation.  The
batch stirrer includes circular cylinders---the stirring rods---that undergo
periodic motion.  The exact velocity field for one circular rod in a Stokes
flow was derived in Ref.~\cite{Finn2001}. For more than one rod there is no
exact expression available for the velocity field, so we use instead a series
expansion suggested by Finn~\etal~\cite{Finn2003}.  We use an adaptive
fourth-order Runge--Kutta integrator for the time-stepping.

\begin{figure}
\begin{center}
\includegraphics[width=7cm,angle=-90]{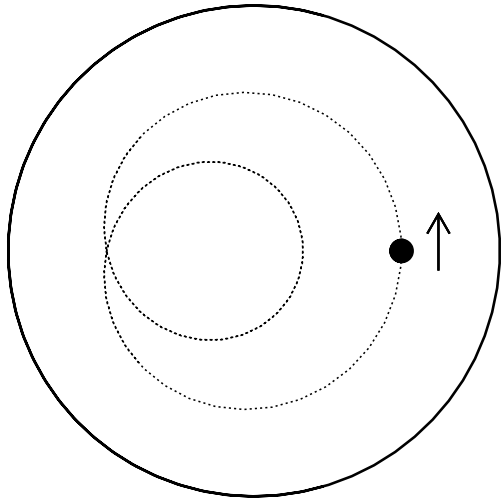}
\end{center} 
\caption{A rod traveling on a epicyclic path. The trajectory of the rod
encloses two different regions, a loop and a crescent.}
\label{fig:trmfig}
\end{figure}

We study first a configuration of the translating rotating mixer (TRM) defined
by Finn~\etal~\cite{Finn2001}.  The system consists of a two-dimensional disk
stirred by a circular rod that moves around in the disk.  The center of the rod
moves on an epicyclic path (in a time period $T$) given by
\begin{equation}
\begin{split}
x(t) &= r_{1} \cos{2\pi m t/T} + r_{2} \cos{2\pi n t/T},\\
y(t) &= r_{1} \sin{2\pi m t/T} + r_{2} \sin{2\pi n t/T},
\end{split}
\label{eq:trmtraj}%
\end{equation}
as shown in Fig. \ref{fig:trmfig}. Note that such a path can be implemented in
a real mixing device using straightforward gearing. It is possible to choose
very complicated trajectories by changing $m$ and $n$, however we limit
ourselves to the comparatively simple case $m=1$ and $n=2$.  The other
parameter values used here are $r_{1}=0.2$, $r_{2}=0.5$.  The radius of the
outer disk is $1$ and we tested configurations with different values for the
rod radius $a_{\mathrm{in}}$.

\begin{figure}
\subfigure[]{
\includegraphics[width=7.25cm]{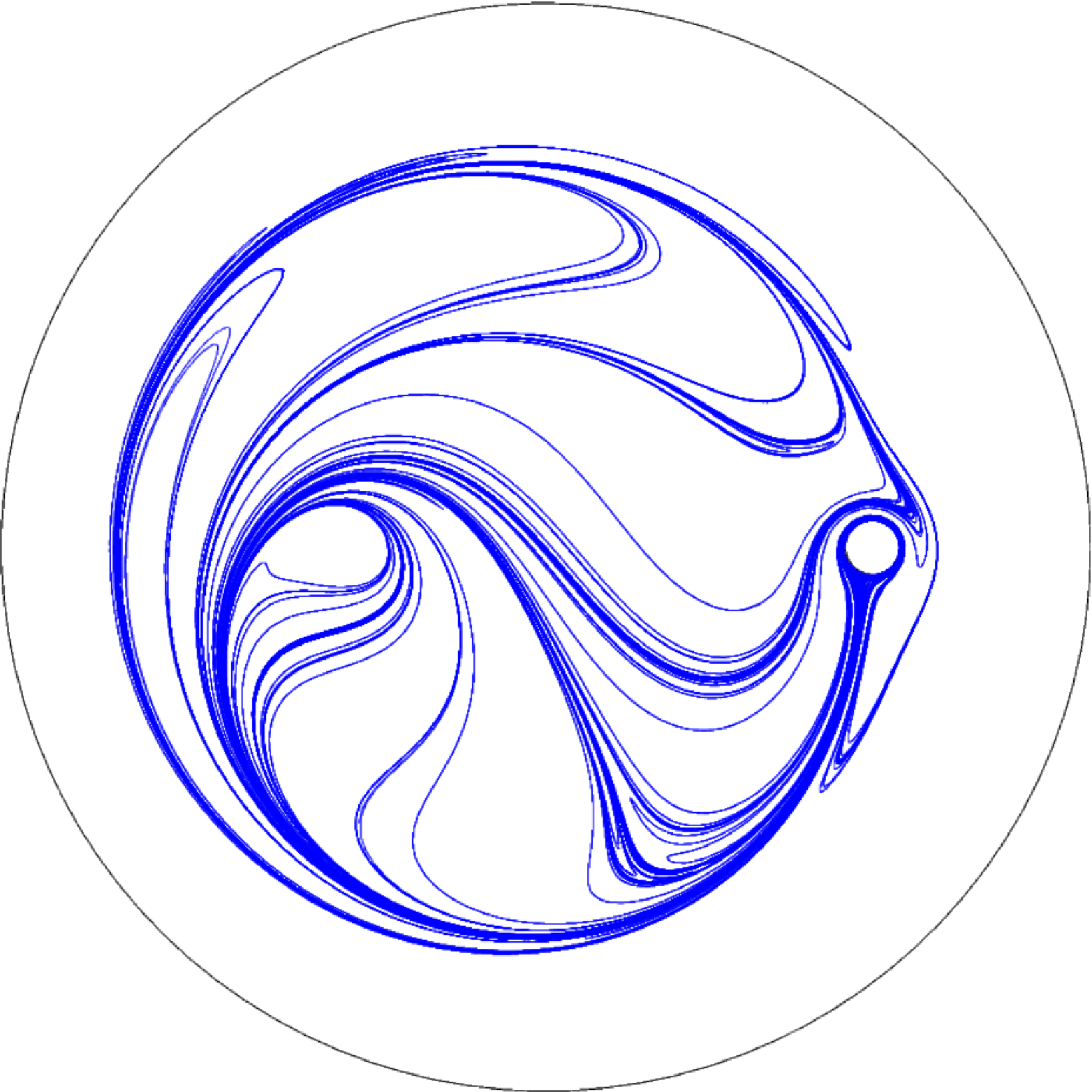}
\label{fig:trmblob}
}\hspace{1cm}
\subfigure[]{
\includegraphics[width=7.25cm]{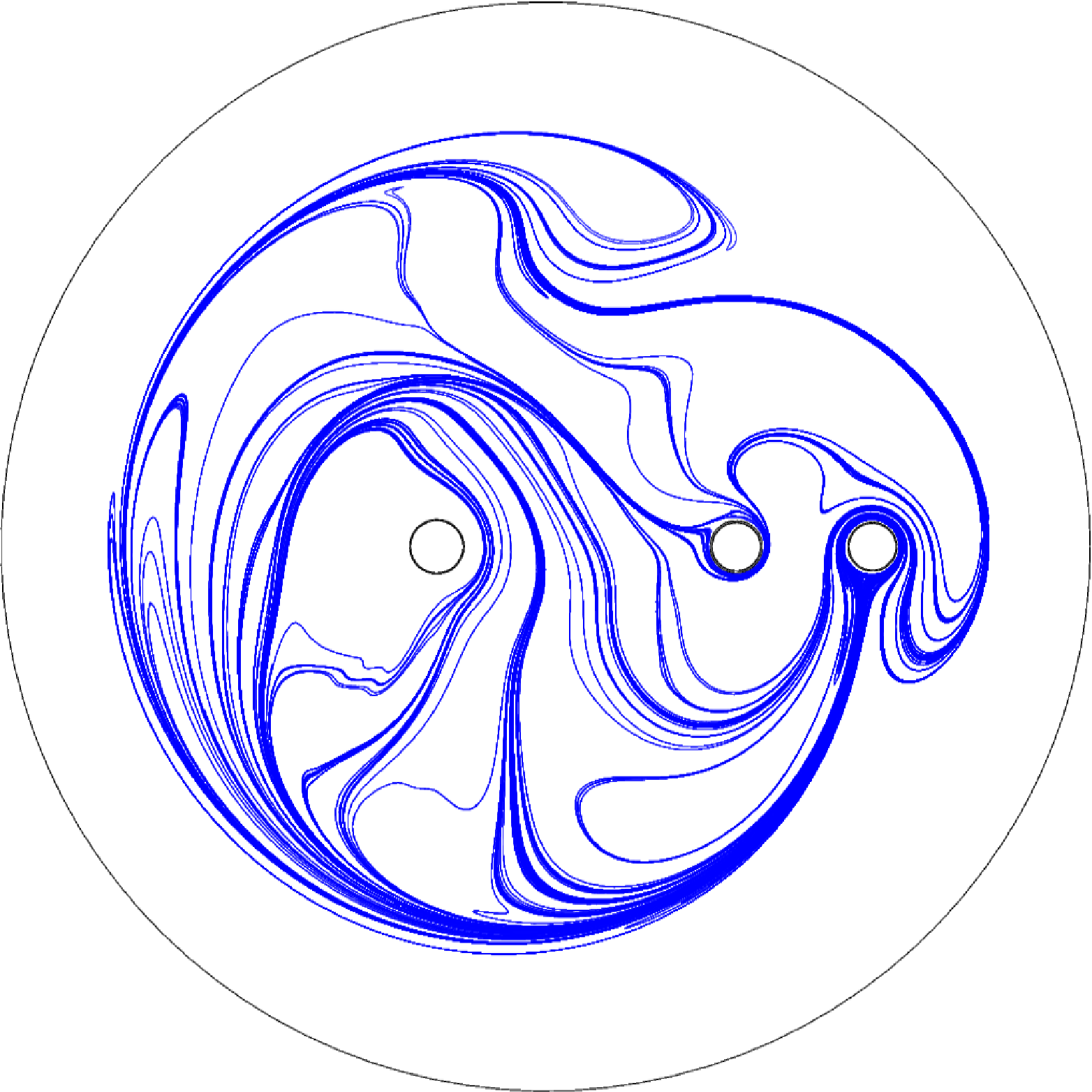}
\label{fig:modiftrmblob}
}
\caption{(a) The stretching of a small blob initially surrounding the moving
rod; (b) Same as in (a), except two extra fixed rods are inserted in the flow.
The patterns made by the blob in (a) and (b) look similar, suggesting that
there are invisible topological obstacles in (a) as the fixed rods in
(b). Note in particular the similarity between the fixed rod inside the loop
in (b) and the way the line wraps itself around an invisible obstacle inside
the loop in (a). }
\label{fig:trmblobab}
\end{figure}

There is only one rod so the one-strand braid formed by the stirrer is
trivial.  Topologically, the motion of this single rod does not imply a
positive lower bound on the topological entropy of the flow.  This does not
mean that material lines cannot grow exponentially for this flow.  We have
plotted in Fig.~\ref{fig:trmblob} the image of a small blob, \ie\ a circle
enclosing the rod at $t=0$, after just four periods of the flow with
$a_{\mathrm{in}}=0.05$.  The small blob has been tremendously stretched,
suggesting exponential growth.

Furthermore, note some similarities with another stirring protocol, where the
rod is moving on the same path but we have added two fixed rods in the regions
enclosed by the moving rod's trajectory (Fig. \ref{fig:modiftrmblob}). For
this protocol the braid formed with the stirrers
is~$\sigma_{-2}\sigma_{-1}\sigma_{-1}\sigma_{-1}\sigma_{-1}\sigma_{-2}$
\cite{Finn2003} with topological entropy~$\htopb=1.76$.  (From now on, we
shall use~$\htopb$ to denote the topological entropy of a braid, which is a
lower bound on the corresponding flow's topological entropy,~$\htopf$: $\htopb
\leq \htopf$.)  Thus we expect the efficient stretching displayed in
Fig.~\ref{fig:modiftrmblob}.  The braid's entropy is a lower bound for the
entropy of the flow, and we indeed measure~$\htopf=2.38$. (We recall that we
compute~$\htopf$ by calculating the exponential rate of growth of material
lines in the chaotic region.)

There is thus a discrepancy in that the motion of the rod in
Fig.~\ref{fig:trmblob} does not account for the observed exponential
stretching of material lines, at a rate given by~$\htopf=2.32$.  An indication
of the source of the missing topological entropy is that this rate is
comparable to the one observed in the protocol of
Fig.~\ref{fig:modiftrmblob},~$\htopf=2.38$, where extra obstacles are present.
We shall see in the following sections that the ``missing'' topological
entropy can be accounted for by looking at periodic orbits in the flow and
their topological effect on material lines.

\subsection{Elliptic Islands as Ghost Rods}

In Section~\ref{sec:trm} we saw that much of the topological entropy in a
one-rod protocol could be accounted for by adding two fixed rods to the flow.
These fixed rods modify the flow, but they do not significantly modify the
topology of advected material lines compared to the single-rod case
(Fig.~\ref{fig:trmblobab}).  Hence, for the TRM with only one rod we observe
that material lines grow \emph{as if} rods were present inside the loops
traced by the physical rod's trajectory.  This justifies introducing the
notion of ``ghost rods'': something inside the physical rod's trajectory is
playing the role of a real rod, and we shall soon see that in this case
elliptical islands are the culprits.  These islands braid with the physical
rod, and taken together they give a positive topological entropy.  In general,
we refer to periodic structures of the flow (islands or isolated points) as
ghost rods when they play a role in determining the topological entropy.
These are topological obstacles and are thus candidates for forming nontrivial
braids.  The topological approach puts all periodic structures---orbits and
rods---on the same footing.

\begin{figure}
\includegraphics[width=0.4\textwidth]{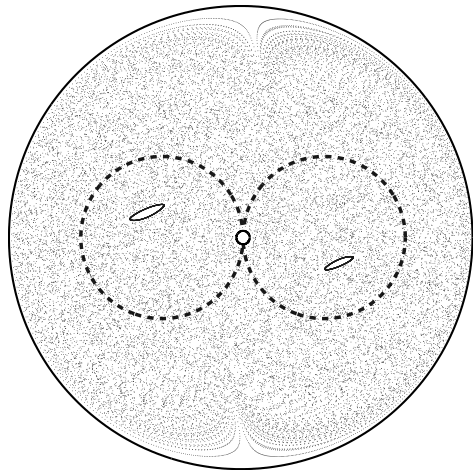}
\caption{Poincar\'e section for a protocol with one rod traveling on a
figure-eight path (dashed line). Two regular islands are present inside
each loop of the rod's path. They are topological obstacles that form
a non-trivial braid with the rod.}
\label{fig:boytrm}
\end{figure}

The introduction of ghost rods becomes even more relevant if one considers the
one-rod protocol pictured in Fig.~\ref{fig:boytrm}.  The rod is moving on a
figure-eight path, traveling clockwise on the left circle of the eight and
anticlockwise on the right circle. A Poincar\'e section reveals two small
islands inside both circles (see Fig.~\ref{fig:boytrm}).  Initial conditions
inside these islands remain there forever and are thus topologically
equivalent to a fixed rod inside each circle of the figure-eight.  By studying
the motion of the rod closely, it is easy to show that the braid formed by the
rod and the islands is~$\sigma_{1}\sigma_{-2}\sigma_{-2}\sigma_{1}$, which has
a topological entropy $\htopb=1.76$.  Indeed, we measure a topological
entropy~$\htopf= 2.25$, which is
greater than $1.76$.  (We shall account for the difference with the measured
topological entropy of the flow later.)  Hence, although elliptic islands are
usually considered barriers to mixing, they can also yield a lower bound on
the topological entropy of the region exterior to them.  All the results
presented here for the figure-eight protocol are for the parameters $a=0.35$
(radius of the circles forming the eight), $a_{\mathrm{in}}=0.04$ (radius of
the rod) and $a_{\mathrm{out}}=1$ (radius of the outer circle).

\begin{figure}
\includegraphics[width=0.4\textwidth]{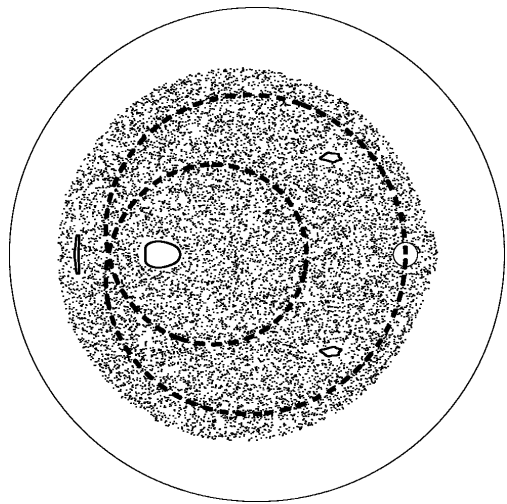}
\caption{Poincar\'e section for the TRM protocol, showing different kinds of
topological obstacles. The physical rod is moving on an epicyclic path whereas
a regular island   stays inside
the loop and three islands
of period three are permuted each period. The braid formed with the rod and
these islands is shown in Fig.  \ref{fig:TRMislandsbraid}}
\label{fig:trmpoinc}
\end{figure}

\begin{figure}
\centering
\subfigure[]{
\begin{minipage}{0.4\textwidth}
\includegraphics[width=0.9\textwidth]{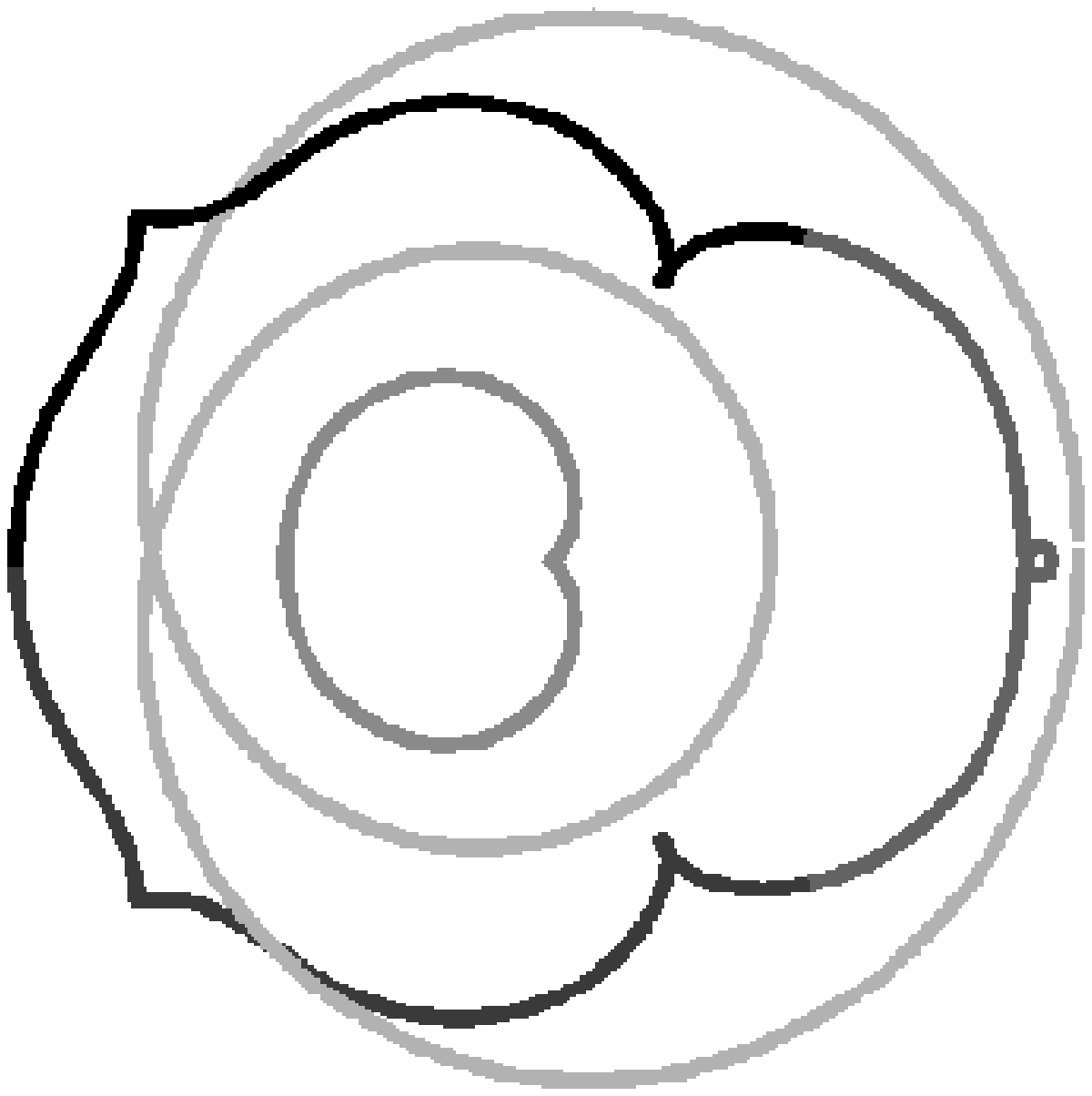}
\end{minipage}
\label{fig:TRMislanda}
}\hspace{1cm}
\subfigure[]{
\begin{minipage}{0.4\textwidth}
\includegraphics[width=\textwidth,height=1.2\textwidth]{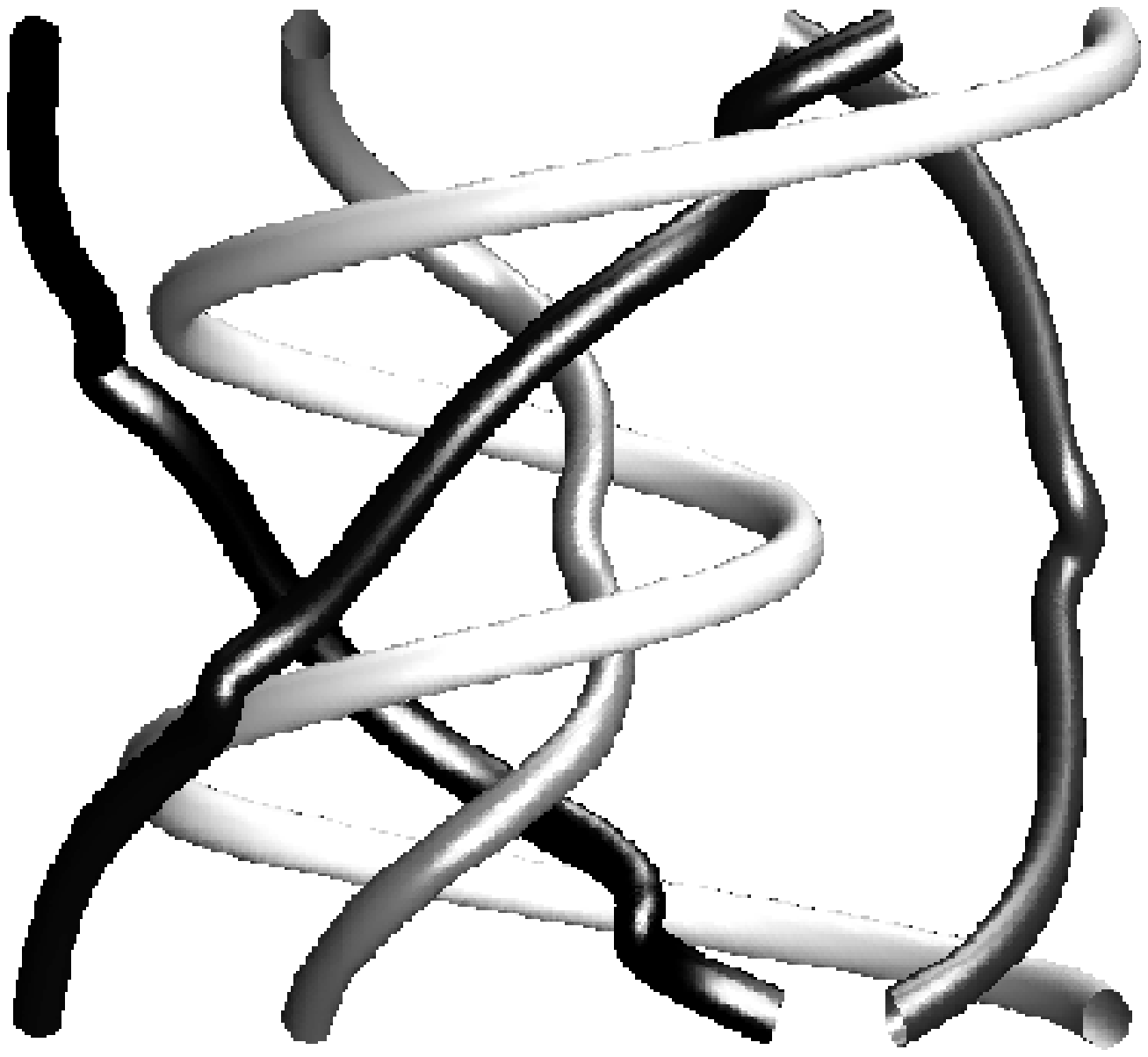}
\end{minipage}
\label{fig:TRMislandb}
}
\hfill
\caption{(a) Trajectories of the topological obstacles (the physical rod and
two periodic islands) shown in Fig.~\ref{fig:trmpoinc}.  (b) The same
trajectories in a space-time diagram form a braid with positive topological
entropy~$\htopb=1.72$.}
\label{fig:TRMislandsbraid}
\end{figure}

So far we have only considered period-1 islands that stay into the regions
bounded by the rod's trajectory, but this is not always possible.  In general
we have to consider more complicated orbits. For instance, the two islands
inside the eight (Fig.~\ref{fig:boytrm}) are not present for protocols with a
larger rod radius~$a_{\mathrm{in}}$: in that case we have not found any points
that remain forever inside one of the circles.  Similarly, for the epicyclic
path (Fig.~\ref{fig:trmfig}) there is an island inside the loop part of the
trajectory; however any point will leave the crescent region after a finite
time because of the ascending movement induced by the rod, so there are no
fixed ghosts rods in that region.  The Poincar\'e section shown in
Fig.~\ref{fig:trmpoinc} suggests however other candidates for ghost
rods. First, as we noted before, there is a period 1 island that remains
inside the loop forever.  Second, the Poincar\'e section reveals three islands
that are part of the same period-3 structure.  Two ``images'' of this period-3
island are inside the crescent.  The braid formed by the rod and these four
islands is shown in Fig.~\ref{fig:TRMislandsbraid}.

We now have a first method of computing a lower bound on the topological
entropy of a flow: look for elliptic islands and calculate the braid formed by
the physical rod(s) (if any) and the islands (the ghost rods). Its
entropy~$\htopb$ will be a lower bound on the topological entropy of the
flow~$\htopf$.  However, this lower bound is often not a very good one: we
show in the next section that it is possible to improve it by considering not
only islands, but also more general (unstable) periodic orbits.

\subsection{Unstable Periodic Orbits as Ghost Rods}
\label{sec:upoghostrods}

A positive topological entropy implies a horseshoe structure~\cite{Katok1980}
and thus an infinite number of unstable periodic orbits (UPO's) in the
flow. These periodic points are topological obstacles as well as the physical
rods (though they have zero size), and hence are ghost rods.  However, most
periodic points are unstable and therefore difficult to detect: a trajectory
initialized near an unstable periodic point will diverge from the periodic
orbit (exponentially for a hyperbolic orbit).  We use the method of Schmelcher
and Diakonos~\cite{Schmelcher1998} to detect periodic orbits numerically.  The
method relies on finding the periodic orbits of a modified version of the flow
that has the same periodic orbits, except that they are stable in the modified
flow.  Diakonos~\etal~\cite{Diakonos1998} point out that this method selects
the least unstable orbits, that is, the least unstable orbits are found first
and one has to change a parameter in the algorithm and therefore increase the
computing time to find more unstable orbits.  Furthermore, we are dealing with
systems with high topological entropy: as asymptotically the flow has
roughly~$\exp{(\htopf n)}$ periodic points of period~$n$, systematic detection
of periodic points is impossible for orbits of high order.  We choose rather
to detect only the least unstable orbits: we show in the next section that it
is possible to derive an accurate value of the topological entropy of a flow
from these orbits.

\section{Calculating the Topological Entropy with Ghost Rods}
\label{sec:braiding}

Boyland~\cite{Boyland1994}, using results by Katok~\cite{Katok1980}, proved
that for an orientation-preserving diffeomorphism of the disk there exists a
sequence of orbits whose entropies converge to the topological entropy of the
flow.  The topological entropy can thus be obtained from the periodic-orbit
structure of the flow.  It should therefore be possible to find a periodic
orbit whose braid has a topological entropy arbitrarily close to the
topological entropy of the flow.  We investigate this before considering
multiple periodic orbits.

\begin{figure} 
\subfigure[]{
\includegraphics{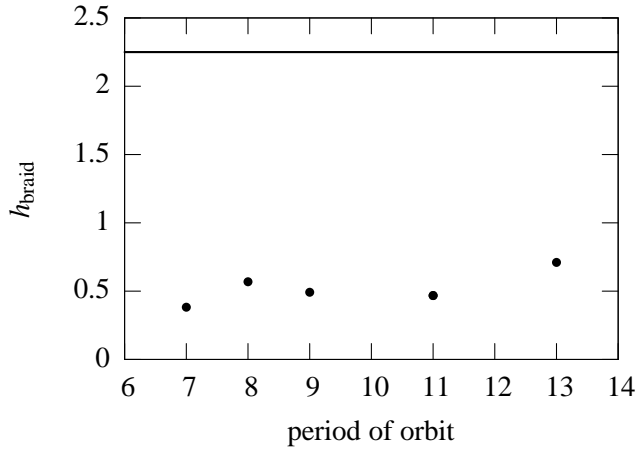}%
\label{fig:self_braiding}
}
\subfigure[]{
\includegraphics{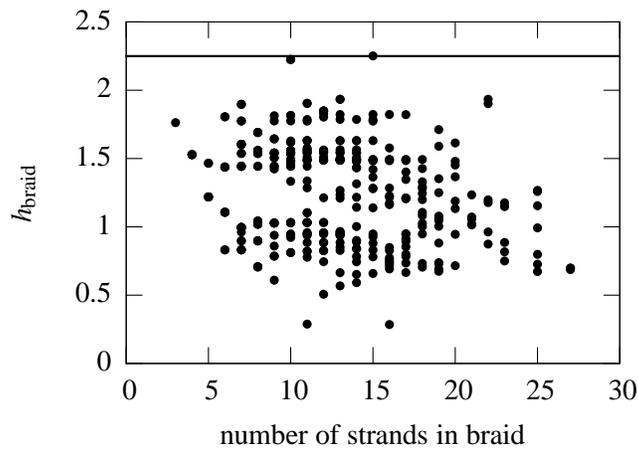}%
\label{fig:braiding_withrod} 
} 
\caption{(a) Some periodic orbits are
self-braiding, that is, the entropy of the braid formed by all the points
of the orbit is strictly positive.  The entropy of the self-braiding
orbits is plotted here vs their period. (b) The positive entropy of
braids formed by the rod's trajectory and pairs of periodic orbits vs the
number of strands in the braid.  In both plots, the solid line is the measured
topological entropy of the flow, $\htopf=2.25$.}
\label{fig:UPOhtop} 
\end{figure}

A periodic orbit is \emph{self-braiding} if the points in the orbit form
a nontrivial braid when taken together.  Figure~\ref{fig:self_braiding}
shows
the topological entropy of self-braiding orbits as a function of their
period for the figure-eight protocol.  Some orbits indeed have a positive
topological entropy, but their entropy is far from the one computed with
the line-growth algorithm ($\htopf=2.25$), solid horizontal line in
Fig.~\ref{fig:self_braiding}.  Furthermore very few orbits self-braid,
\ie\ have a positive entropy.  Hence, it appears that looking at the
topological entropy of individual orbits is not very useful: obtaining a
reasonable approximation to the topological entropy requires orbits of
prohibitively high order or instability.

As self-braiding orbits do not seem very convenient to approach~$\htopf$ we
choose rather to combine several orbits together to form more complex
braids. We first combine each periodic orbit with the rod and we obtain higher
values for~$\htopb$, although still far from~$\htopf$. We also consider the
braids formed by the rod and pairs of periodic orbits, which are also
invariant sets of the first return map.  We consider pairs of orbits because
this is the minimal number of orbits that can capture the topology determined
by the rod's trajectory.  As is shown in Fig.~\ref{fig:braiding_withrod}, we
obtain braid entropies far closer to the value measured for the flow: some
braids approximate the measured entropy to within numerical error. It is thus
more efficient to consider combinations of orbits rather than only
self-braiding orbits: one would surely need orbits of very high order, or very
unstable, to get as close to the measured value~$\htopf$.  For this example we
used a set of $52$ periodic orbits whose positive Floquet exponent is smaller
than $3/T$, where $T$ is the period of the orbit.  Note that although we did
not detect all or even a large number of periodic orbits (there is an infinity
of them), our combination of ghost rods provides us with a very good
approximation of the topological entropy of the flow. For a given system the
entropy lower bound must increase as more orbits are added to a given braid;
nevertheless we see that for the system we consider, only a small number of
orbits is needed to obtain a satisfactory lower bound on~$\htopf$.  We
therefore suggest an alternative method for calculating the topological
entropy of a flow: (i) detect some periodic orbits; (ii) calculate the maximum
entropy of all the braids one can create with these points and the rod(s); and
(iii) see if this maximum entropy converges when one increases the number of
periodic orbits.

\section{Discussion}
\label{sec:discussion}

To summarize, we have characterized chaos in two-dimensional
time-periodic flows by considering the braids formed by periodic points
and calculating their topological entropy for different stirring
protocols.  We have demonstrated the role of periodic points in fluid mixing,
and called these periodic points ghost rods because their movement
stretches material lines as real stirring rods do. This work is an
extension of the topological kinematics theory introduced by
Boyland~\etal~\cite{Boyland1994,Boyland2000,Boyland2003}, since it
characterizes the mixing in a flow by studying the topological constraint
induced by the ghost rods and not only stirrers.  We expect this approach
to develop further in the near future, and to yield new insight on
efficient mixing devices.

The idea of characterizing dynamics of homeomorphisms of surfaces by
puncturing at periodic orbits dates back to Bowen~\cite{Bowen1978}, and the
study of braids formed with periodic points had already been suggested by
Boyland for the general study of diffeomorphisms of the
disk~\cite{Boyland1994}.  In addition, this technique is used in other fields
such as the study of optical parametric oscillators~\cite{Amon2004}.  However,
the present work is to our knowledge the first study of ghost rods in fluid
mechanics.

\begin{figure}
\begin{center}
\includegraphics[width=0.5\textwidth]{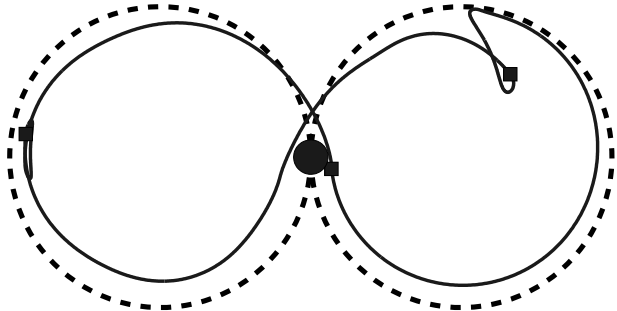}
\end{center}
\caption{The trajectory of the rod (dashed line) and three period-3 points
belonging to the same orbit (solid line). The
position of the rod (filled circle) and the periodic points (filled squares)
at~$t=0$ is also shown.}
\label{fig:boycircle_orbits}
\end{figure}

In contrast to other applications, in our systems the ghost rods are created
by the movement of the physical rod, so we may hope to derive some information
about ghost rods from the motion of that physical rod.  For instance, let us
consider the period-3 orbit for the figure-eight protocol shown in
Fig.~\ref{fig:boycircle_orbits}
(this orbit is more unstable than the ones we used in the previous
section). One point of the orbit is located very close to the physical rod at
$t=0$. Its trajectory will thus be very close to the rod's trajectory at the
beginning of the period.  Later the rod leaves the periodic point in its trail
after a time equal to about~$T/3$.  One period later the rod drags this point
again on its trajectory as it comes close to it.  This accounts for the
topological similarity between the trajectory of the rod and the periodic
orbit.  The braid formed by these period-3 orbits is~$\sigma_{-1}\sigma_{2}$,
which is also the braid studied by Boyland~\etal\ in~\cite{Boyland2000}.  It
is actually the braid with the maximal $\htopb$ that we can form with points
moving on a trajectory strictly equivalent to the rod's path. Indeed we cannot
form a braid with fewer than three points, and periodic points with a higher
period on such a path move slower as they take longer to cover the whole path,
so they have a less efficient braiding (fewer exchanges per period). We
conjecture that this type of figure-eight trajectory is characteristic of all
mixing protocols with a rod moving on a figure-eight path.

We have indeed found such a braid---formed by a period-3 orbit equivalent to
the rod's trajectory---for all the figure-eight protocols we have tested.  We
have tried the protocol plotted in Fig.~\ref{fig:boytrm} with different radii
for the rod, as well as a protocol with one rod moving on a lemniscate (a
lemniscate is the more natural ``figure-eight'' shape~\cite{lemniscate}), and
we have detected this figure-eight periodic orbit associated with the
$\sigma_{-1}\sigma_{2}$ braid for each protocol.  It is thus very tempting to
conjecture that for all the Stokes flows created by the movement of a rod on a
figure-eight path, the topological entropy of the flow is greater than the
entropy of the braid $\sigma_{-1}\sigma_{2}$, that is, $0.96$.  If this
conjecture holds, then ghost rods can be used not only for diagnosing chaos
and calculating the entropy of a flow, but also topological arguments can be
used to \emph{predict} a minimum entropy just from the path of the physical
rod with ghost rods in its trail.  This is a generalization to ghost rods of
the arguments used by Boyland~\etal~\cite{Boyland2000} to predict a universal
minimum entropy from the movements of the rods for three or more physical
rods.  A further natural question then is which ghost rods have the best
braid, that is, the braid giving the topological entropy of the flow, and
could we relate these ``most efficient'' ghost rods to the physical properties
of the flow?  This will be the topic of future research.

The ghost rods approach should also be compared with recent work on
nonperiodic points.  Thiffeault~\cite{Thiffeault2005} noticed that every fluid
particle in a two-dimensional flow is a topological obstacle, much like a
stirrer, and calculated entropies of braids formed by arbitrary chaotic
orbits.  As these points were not periodic points these entropies may not give
a lower bound on the entropy of the flow for short times; however, they can
yield finite-time information about the stretching rate of material
lines. Furthermore, if one considers long time series, a randomly chosen point
in an ergodic chaotic region will repeatedly come very close to periodic
orbits.  It should thus be possible to observe braids with similar properties
as the ones formed by ghost rods.

Finally, the study of ghost rods can be used to prove that a map is chaotic in
the case where some periodic points can be calculated analytically, as for the
sine flow map~\cite{Liu1994a,Finn2005} or the closely-related standard
map~\cite{Chirikov1996}.  Finn~\etal~\cite{Finn2005}, for example, have shown
that the sine flow map has chaotic trajectories for some parameter values.

\section*{Acknowledgments}

We thank Toby Hall for use of his train-tracks code.  This work was funded by
the UK Engineering and Physical Sciences Research Council grant GR/S72931/01.


\end{document}